\documentclass[a4paper,12pt,oneside]{article}
\usepackage{graphicx}
\usepackage{amsmath}
\usepackage{amssymb}
\usepackage{euscript}

\begin{document}

\title{
Interpretation of the Veiling of the Photospheric Spectrum
for T Tauri Stars in Terms of an Accretion Model.
}

\author{A. V. Dodin and S. A. Lamzin }

\date{ \it \small
Sternberg Astronomical Institute, Universitetskii pr. 13, Moscow, 119992 Russia\\
e-mail: dodin\_nv@mail.ru
}

\maketitle



\medskip




\section*{Abstract}

The problem on heating the atmospheres of T Tauri stars by radiation from an accretion shock
has been solved. The structure and radiation spectrum of the emerging so-called hot spot have been
calculated in the LTE approximation. The emission not only in continuum but also in lines has been taken
into account for the first time when calculating the spot spectrum. Comparison with observations has
shown that the strongest of these lines manifest themselves as narrow components of helium and metal
emission lines, while the weaker ones decrease significantly the depth of photospheric absorption lines,
although until now, this effect has been thought to be due to the emission continuum alone. The veiling
by lines changes the depth of different photospheric lines to a very different degree even within a narrow
spectral range. Therefore, the nonmonotonic wavelength dependence of the degree of veiling $r$ found for
some CTTS does not suggest a nontrivial spectral energy distribution of the veiling continuum. In general,
it makes sense to specify the degree of veiling $r$ only by providing the set of photospheric lines from which
this quantity was determined. We show that taking into account the contribution of lines to the veiling
of the photospheric spectrum can cause the existing estimates of the accretion rate onto T Tauri stars to
decrease by several times, with this being also true for stars with a comparatively weakly veiled spectrum.
Neglecting the contribution of lines to the veiling can also lead to appreciable errors in determining the
effective temperature, interstellar extinction, radial velocity, and $v\,\sin i.$


\section*{Introduction}

  Long ago Joy (1949) noticed that the depths and
equivalent widths of photospheric lines in the spectra
of T Tauri stars were smaller than those for main-sequence
stars of the same spectral types, especially
at short wavelengths. This effect is commonly explained
by the fact that the absorption lines of the
stellar photosphere are "veiled"{} by the emission continuum,
the understanding about the nature of which
changed as the views of the cause of activity in T Tauri
stars changed.

  The emission in optical lines and continuum as
well as the very intense ultraviolet (UV) and X-ray
emissions had long been thought to be due to the
existence of thick chromospheres and coronas around
young $(t < 10^7$ yr) low-mass $(M<2\,M_\odot)$ stars that
to some extent are analogous to the solar ones (see
the reviews by Bertout (1989) and references therein). However,
it had become clear by the early 1990s that
this explanation is appropriate only for moderately active
young stars in the spectra of which the equivalent width $(EW)$
of the $H_\alpha$ emission line does not exceed 5-10 \AA\,
there is virtually no veiling. Below, we will not talk about
these objects called weak-lined T Tauri stars.

  Here, we will deal with the so-called classical
T Tauri stars $(EW_{H\alpha} > 10$ \AA). Their observed manifestations
can be explained in terms of the model
of mass accretion from a protoplanetary disk onto
a young star that has a global magnetic field with
a strength $\sim 1-3$ kG. The model suggests that the
matter from the inner disk is frozen in the magnetic
field lines and slides down toward the star along
them, being accelerated by gravity to a velocity $V_0 \sim 300$ km\,s$^{-1}$. A shock wave at the front of which the gas velocity decreases approximately by a factor of 4
and the gas is abruptly heated to a temperature of $\sim 10^6$ K emerges near the stellar surface.
The postshock matter cools down while gradually radiating its
thermal energy in the UV and X-ray ranges and settles
to the stellar surface while reducing its velocity.

  One half of the short-wavelength radiation flux
of the shock from the cooling zone escapes upward,
heating and ionizing the pre-shock gas, and the second
half irradiates the star, producing the so-called
hot spot on its surface. Estimations (K{\"o}nigl 1991;
Lamzin 1995) and numerical calculations (Calvet and
Gullbring 1998) show that for classical T Tauri stars
(CTTS) at a particle number density in the gas onflowing
onto the front $N_0$ above $\sim 10^{13}$ cm$^{-3}$, the
pre-shock region becomes opaque in optical continuum.
This means that the shock photosphere must
be located upstream of its front at $\log N_0 > 13$ and
downstream of its front at lower $N_0$.

  Now, there is no doubt that the so-called narrow
components of emission lines in the spectra of CTTS
are formed inside the hot spot (see Dodin et al. (2012)
and references therein). The observability of these
components implies that the shock photosphere is
in even deeper layers, i.e., in the hot spot and not
upstream of the front. Since half of the kinetic energy
of the accreting material is radiated in the spot, it is
natural to assume that precisely the hot-spot photosphere
is the source of the veiling continuum.

   If {\it all} the short-wavelength radiation of the shock
incident on the stellar atmosphere is assumed to be
reemitted outward in the form of an emission continuum,
then the mass accretion rate onto the star
$\dot M$
can be found from the relation $L_c =
0.5\cdot \dot M V_0^2/2,$
where  $L_c$ is the bolometric luminosity of the veiling
continuum, which, just as the pre-shock gas velocity,
can be determined by analyzing the spectrum. Basically,
the $\dot M$ estimates for most CTTS were obtained
precisely in this way (see, e.g., Valenti et al. 1993;
Hartigan et al. 1995; Gullbring et al. 1998, 2000).

   The only (at present) calculation of the vertical
hot-spot structure and the veiling continuum spectrum
was performed by Calvet and Gullbring (1998)
without any allowance for the emission in lines. Comparison
of the calculated and observed spectra allowed
one not only to determine the accretion rate and
the hot-spot sizes but also to self-consistently find
the star's spectral type and the interstellar reddening,
which are used to determine the emission continuum
spectrum from observations. Moreover, as yet there
is no other way to reliably determine the spectral type
and the degree of interstellar reddening precisely for
heavily veiled CTTS.

   A method for separating the veiling continuum
by comparing the equivalent widths of photospheric
lines in the spectra of CTTS and a comparison star
was proposed by Hartigan et al. (1989) and has been
used with slight modifications up until now. Having
analyzed the spectrum of the star BP Tau, Hartigan
et al. (1989) concluded that the veiling was attributable
precisely to the emission continuum rather
than stemmed from the fact that weak emission lines
were superimposed on photospheric lines, thereby
reducing their depth.

   However, Petrov et al. (2001) found that the
presence of emission lines inside photospheric ones
in the spectrum of RW Aur led to noticeable observed
effects, while Gahm et al. (2008) and Petrov
et al. (2011) showed that for several heavily veiled
CTTS, the emission in lines contributed significantly
to the decrease in the depth of photospheric lines.
Theoretically, the presence of emission lines in the
hot-spot radiation spectrum seems quite natural,
because the temperature above the spot photosphere
increases outward. It is reasonable to assume that the
strongest of these lines manifest themselves in the
spectra of CTTS as narrow emission components,
while the weaker ones to a certain extent blend the
photospheric lines.

   Since the contribution of lines to the veiling has
been disregarded so far, it can be concluded that the
intensity of the emission continuum in the spectra
of CTTS has been systematically overestimated and,
hence, all of the available calculated accretion rates
have also been overestimated. The goal of this paper
is to calculate the radiation spectrum of the hot spot
not only in continuum but also in lines and to apply
the results obtained to ascertain the extent to which
allowance for the emission in lines can change the
available estimates of the accretion rate and effective
temperature of CTTS and the interstellar extinction.


\section*{Formulation of the problem and input parameters}

   To calculate the hot-spot radiation spectrum, the
problem on heating the atmosphere of a young star by
radiation from an accretion shock should be solved.
The heating of a stellar atmosphere by external radiation
has been studied in many papers devoted to the
reflection effect in binary systems (see the monograph
by Sakhibullin (1997) and references therein). However,
these calculations cannot be directly used to
determine the radiation spectrum of the hot spots on
CTTS for the following reasons. First, the radiations
from the hot companions of stars and the accretion
shock are different in spectral composition. Second,
in our case, the atmosphere being heated is immediately
adjacent to the region that serves as an irradiation
source. Consequently, the radiation from all sides
will be incident on each point of the hot spot, while
the radiation from the hot companion arrives at each
point of the atmosphere of the neighboring star in the
form of an almost parallel flux. For the same reason,
in our case, the pressure at the outer boundary of the
atmosphere being heated should be equal not to zero
but to the pressure that is established far downstream
of the shock front (Zel'dovich and Raizer 1966):
\begin{equation}
P_0 =  \rho_0 V_0^2,
 \label{P0-exp}
\end{equation}
where $\rho_0 = 1.3 m_p N_0$ is the density of the pre-shock
gas (with solar elemental abundances).

     Suppose that the atmosphere being heated is
stationary and consists of plane-parallel layers of
gas with solar elemental abundances. We take the
radiation spectrum of the post-shock region from
Lamzin (1998), where the problem on the shock
structure was solved under similar assumptions. If
the influence of gravity on the gas motion is disregarded,
then the accretion shock structure in the case
of CTTS is almost uniquely determined by the preshock
gas density $N_0$ and velocity $V_0$. In this case, the
radiation spectrum of the post-shock region depends
mainly on $V_0:$ the greater the latter, the higher the
maximum post-shock temperature, and the harder
the spectrum. The geometrical sizes of the pre-shock
heating region and the post-shock cooling region as
well as $P_0$ in Eq. (\ref{P0-exp}) depend on $N_0.$

   The radiation from both the post-shock region
and the pre-shock zone is incident on the stellar
atmosphere. The radiation from the post-shock zone
consists almost entirely of photons with energies from
5 eV to 1 keV. To reduce the computational time, this
range in Lamzin (1998) was divided into several tens
of energy intervals each of which was considered as
a pseudo-line with a frequency $\nu_i,$ equal to the mean
frequency inside the interval and with a flux $F_i$ equal
to the total flux of the actual lines and continuum
falling into this interval. The spectral flux density $F_\nu^i$
needed for our calculations was obtained by dividing
$F_i$ by the width $\Delta \nu_i$ of the corresponding interval.

   Since the radiation spectrum emergent from the
pre-shock zone was not calculated by Lamzin (1998),
we calculated it separately by the technique described
in the Appendix. The material in this zone not only
radiates toward the star but also absorbs the radiation
emergent from the hot spot, changing its spectrum.

   Taking into account the gravitational potential of
CTTS, we will use $V_0$ in the range from 200 to
400 km\,$s^{-1}$. As regards the range of $N_0,$ in accordance
with what was said in the Introduction, we will
take $\log N_0=13$ and $\log N_0=11$ as the upper and
lower boundaries of the range, respectively, because,
as we will see below, the manifestation of accretion
will be essentially unnoticeable at lower densities.
Apart from the parameters $V_0$ and $N_0$, to calculate the
vertical structure and radiation spectrum of the hot
spot, we should additionally specify the star's effective
temperature $T_{eff}$, surface gravity g, and microturbulence
$V_t$ in the atmosphere being heated. In our
calculations, we varied $T_{eff}$ in the range from 3750 to
5000 K but always set $\log g = 4.0$ and $V_t = 2$ km s$^{-1}$
to reduce the number of free parameters.


\section*{The method of calculating the vertical hot-spot structure}
  To calculate the structure of an atmosphere being
heated by external radiation, we used the freely
available \texttt{ATLAS9} code (Kurucz 1993; Sbordone
et al. 2004; Castelli and Kurucz 2004) into which
the changes specified below were made. A detailed
description of the code is given in Kurucz (1970),
and we will only point out some of its peculiarities
important for our problem.

  The original version of the \texttt{ATLAS9} code computes
stationary plane-parallel LTE models of hydrostatically
equilibrium atmospheres with a constant
(in depth) energy flux transferred by radiation and
convection. The equations describing the mechanical
equilibrium condition and radiative transfer are
$$
{ {\rm d}P \over {\rm d}z } = -g \rho, \qquad
{ {\rm d}\tau_\nu \over {\rm d}z } = - \rho \chi_\nu.
$$
$$
\mu { {\rm d}I_\nu \over {\rm d}\tau_\nu } = I_\nu - S_\nu,
\qquad
S_\nu = \left( 1-\alpha_{\nu} \right){B_\nu} + \alpha_{\nu}J_{\nu}.
$$
Here, $P$ is the gas pressure;  $\tau_\nu$ and $\chi_\nu$ are the optical
depth and the (lines + continuum) absorption
coefficient at frequency $\nu$, respectively; $I_\nu (\tau_\nu)$ is the
radiation intensity, $S_\nu = S_\nu (\tau_\nu)$ is the source function,
$J_\nu (\tau_\nu)= \int I_\nu \, {\rm d}\Omega /4\pi$ is the radiation intensity
averaged over a solid angle $\Omega$; $\alpha_\nu$ is the ratio of the
scattering coefficient to the total (lines + continuum)
absorption coefficient; and $B_\nu = B_\nu(T)$ is the Planck
function.

  The relation expressing the energy conservation
law is
\begin{equation}
H_{conv}(\tau_{Ross}) + H_{rad}(\tau_{Ross})
= \frac{\sigma T_{eff}^4}{4\pi},
 \label{H-const-0}
\end{equation}
where $H_{conv}$ and $H_{rad}$  are the heat fluxes at the Rosseland
optical depth $\tau_{Ross}$ transferred by convection and
radiation, respectively. Convection is described in the
code in terms of the mixing-length theory. In our
calculations, we took the ratio of the mixing length
to the pressure scale height to be $l/H_p = 1.25.$

The blanketing effect is taken into account in the
code by the so-called ODF(opacity distribution function)
method for 337 frequency intervals in the range
of wavelengths $\lambda$ from 91 {\AA} (136 eV) to 160 $\mu$m.
Using interpolation, we recalculated the shock radiation
spectrum from Lamzin (1998) to these frequency
intervals. We did not consider the transfer of radiation
with $\lambda < 91$ \AA{}, adding the shock radiation energy in
this range to the frequency cell with $\lambda = 91$ \AA, lest
the original code be changed significantly. We believe
that such a simplification does not lead to a significant
error, because less than 10\% of the bolometric luminosity
of the shock is concentrated in the region with
$\lambda < 91$ \AA{} even at $V_0 = 400$ km\,s$^{-1}.$

To take into account the external radiation, we
should know how its intensity $I_{\nu}^0$ changes with the
angle of incidence $\theta$ on the (plane) atmosphere or,
more precisely, the function $I_{\nu}^0(\mu),$ where $\mu = \cos \theta.$
Let us assume that the normal to the atmosphere's
surface is directed outward. Then, $\mu <0$ and $\mu > 0$
will correspond to the radiation incident on the atmosphere
from the outside and the radiation emerging
from the atmosphere outward, respectively.

   It followed from the calculations by Lamzin (1998)
and Calvet and Gullbring (1998) that the optical
depth of the post-shock zone in a direction perpendicular
to the shock front $\tau_{sh}^0 \ll 1$ at all frequencies. On
this basis, these authors assumed that the post-shock
zone was completely transparent to the radiation from
the hot spot and the pre-shock region, while the
radiation from the cooling zone escapes in the form of
two equal (in magnitude) fluxes $F_\nu^{sh},$ one of which is
directed toward the star and the other is directed away
from the star. For convenience, we will assume below
that $F_\nu^{sh}>0.$ The expression for $I_{\nu}^0(\mu)$ can then be
written as
\begin{equation}
I_{\nu}^0(\mu)={ I_{\nu}^{sh}(-1) + I_{\nu}^{pre}(-1) \over {-\mu} }
\, , \qquad
I_{\nu}^{sh} (-1) = { F_\nu^{sh} \over 2\pi }.
  \label{I-thin}
\end{equation}
Here, $I_{\nu}^{pre}$ is the intensity of the radiation coming
from the pre-shock region (see the Appendix) and the
second relation can be derived from the condition
$$
F_\nu^{sh} = -\int^{2\pi}_0 {\rm d}\varphi
\int^0_{-1} { I_{\nu}^{sh} (-1) \over {-\mu} } \, \mu \,{\rm d}\mu =
2\pi I_{\nu}^{sh}(-1),
$$
The minus implies that, by definition, $F_\nu^{sh}>0.$

   However, a plane layer of an infinite extent cannot
be optically thin in all directions: the optical depth
$\tau_{sh}^0$ will exceed one for directions with $|\mu| < \tau_{sh}^0$
no matter how small $\tau_{sh}^0$ is. In reality, the maximum
value of $\tau_{sh}$ is limited by the finite sizes of the hot spot
and/or the curvature of the stellar surface. It is important
to note that properly allowing for these factors
when calculating the shock structure and radiation
spectrum is a nontrivial problem, because we should
have simultaneously taken into account the possibility
of photon escape through the side walls of the
accretion stream and the change in $N_0$ and $V_0$ across
the accretion column, which, in turn, depends on the
geometry of the star's magnetic field. In this case, we
would have to solve a three-dimensional problem of
radiation hydrodynamics with a large number of free
parameters.

   We proceeded as follows. Initially, we calculated
the vertical spot structure by using dependence (\ref{I-thin})
and assuming that $|\mu| \geqslant 0.01.$ Subsequently, we
repeated our calculations by assuming the intensity
of the radiation irradiating the stellar atmosphere to
be independent of the direction:
\begin{equation}
I_{\nu}^0(\mu) = const = { F_\nu^{sh} \over \pi } + I_\nu^{pre},
  \label{I-const}
\end{equation}
where $I_{\nu}^{pre}$ is the (isotropic) radiation intensity from
the pre-shock region.

    An increase in $\tau_{sh}$ with decreasing $|\mu|$ must be
accompanied by a decrease in the intensity of the
emergent radiation as $\theta \rightarrow \pi/2$ and, thus, in the degree
of its anisotropy compared to an optically thin
layer. Therefore, it would be reasonable to expect the
dependence $I_{\nu}^0(\mu)$ in the real situation to be between
the first and second cases. Running ahead, we will
say that the differences in vertical spot structure for
dependences (\ref{I-thin}) and (\ref{I-const}) turned out to be comparatively
small.

   To take the external radiation into account, in the
\texttt{JOSH}  procedure of the \texttt{ATLAS9} code we added the
corresponding terms to the right-hand sides of the
expressions for the mean intensity $J_\nu$ and flux $H_\nu.$
They took the form
\begin{equation}
J_\nu (\tau_\nu) = {1 \over 2}\,
\int_0^\infty  S_\nu (t)\, E_1\left( |\tau_\nu - t |\right) {\rm d}t
+ J_\nu^0 (\tau_\nu)
\equiv \Lambda \left[ S\right] + J_\nu^0,
 \label{J-new}
\end{equation}
$$
H_\nu (\tau_{\nu}) \equiv
\int_{4\pi} \mu I_\nu\, { {\rm d}\Omega \over 4\pi} =
{1 \over 2}\, \int_0^\infty S_\nu(t) E_2 \left( |\tau_\nu - t | \right)\,
{\rm d}t
+ H_\nu^0(\tau_\nu),
$$
where $E_n(x) = \int_1^\infty e^{-xt}\,{\rm d}t/t^n$ is the exponential integral.
The second equality in (\ref{J-new}) is the standard form
of writing this relation as the so-called $\Lambda$-operator.

 In the case of an optically thin layer, the expressions
for $J_\nu^0$ and $H_\nu^0$ are
$$
J_\nu^0 = { F_\nu^{sh} + \pi I_{\nu}^{pre} \over 4\pi } E_1(\tau_\nu),
\qquad
H_\nu^0 = -{F_\nu^{sh} + \pi I_{\nu}^{pre} \over 4\pi } E_2(\tau_\nu),
$$
for the isotropic case,
$$
J_\nu^0 = { F_\nu^{sh}+\pi I_\nu^{pre} \over 2\pi }\, E_2(\tau_{\nu}),
\qquad
H_\nu^0 = -\, { F_\nu^{sh}+\pi I_\nu^{pre} \over 2\pi }\, E_3(\tau_{\nu}).
$$

  We also modified Eq. (\ref{H-const-0}): the term
$\int_0^\infty |H_\nu^0| \,{\rm d}\nu$ in
which the absolute value sign implies that $H_\nu^0 < 0$
was added to its right-hand side. Note, incidentally,
that $H_{rad}$ in Eq. (\ref{H-const-0}) is the flux $H_\nu$ integrated over all
frequencies.

   The method of solving the equations describing
the structure and radiation field of the atmosphere
can be briefly described as follows. A discrete mesh
of optical depths is introduced and the differential
and integral equations at the mesh points $\tau_{\nu i}$ $(i=1,2,...,N)$
 are replaced by algebraic ones, which are
then solved by the method of successive approximations
(Kurucz 1970). In particular, the $\Lambda$-operator
from Eq. (\ref{J-new}) is described by a square matrix $N\times N$
with elements  $\Lambda_{ij}.$ To take the external radiation into
account, we replaced the original relation describing
the process of the so-called $\Lambda$-iterations in the \texttt{ATLAS9}
code by
$$
S_{\nu i} = S^{old}_{\nu i} +
{
\left( 1-\alpha_{\nu i} \right) B_{\nu i} +
\alpha_{\nu i}\, {\Lambda \left[ S^{old}_\nu \right]}_i +
\alpha_{\nu i} \, J_{\nu i}^0 - S^{old}_{\nu i}
\over 1-\alpha_{\nu i}\, \Lambda_{ii}
},
$$
where $S^{old}_\nu$ is the source function from the preceding
iteration.

   We also made changes to the \texttt{TCORR} procedure
of the \texttt{ATLAS9} code, which realizes a temperature
correction within the iteration process: following
the recommendation by Sakhibullin and Shimanskii
(1996), we applied the so-called $\Lambda$-correction
instead of the original algorithm at small $\tau_\nu:$
$$
\Delta T =
\frac
{
\int\limits_0^\infty {\left( J_\nu - B_\nu \right) \chi_\nu } \,
{\rm d}\nu
}
{ \int\limits_0^\infty {\frac{ {\rm d}B_\nu}{{\rm d}T}} \chi_\nu \,
{\rm d}\nu
},
$$
where $\chi_\nu$ is the total absorption coefficient.

    In the case of models with a poor convergence of
iterations, we gradually reduced the maximum temperature
correction, which eventually allowed a satisfactory
solution to be obtained. However, the number
of iterations could reach almost 1000 in this case.

   Finishing the description of the changes made into
the \texttt{ATLAS9} code, recall that at $\tau_\nu=0$ we assumed
the gas pressure $P_0$ to be equal to the value defined
by Eq. (\ref{P0-exp}), not to zero, as in the original version of the
code. Since we will be interested in $\log N_0 \geqslant 11.0$ and
 $V_0 \geqslant 200$ km s$^{-1}$, $\lg P_0 \geqslant 2.$

\begin{figure}
\begin{center}
\includegraphics[scale=0.5]{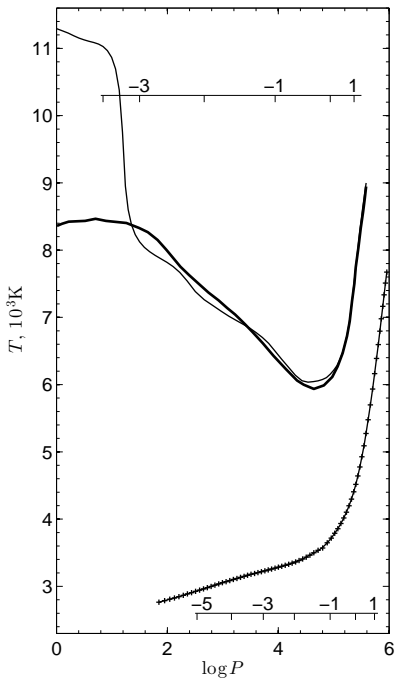}
\end{center}
\caption{
Temperature distribution in a stellar atmosphere
with $T_{eff}=4100$ K and $\log g=4.5$ as a function of pressure
$P$ (dyn cm$^{-2}$). The structure of an atmosphere
without external heating is shown in the lower part of the
figure: the thin solid line indicates the calculation with the
unmodified version of \texttt{ATLAS9}; the crosses represent our
calculations. The scale under the curve is the logarithm
of the Rosseland optical depth for these models. The
structure of the same atmosphere but heated by blackbody
radiation with $T_B=20\,000$ K is shown in the upper
part of the figure: the thick and thin solid lines indicate
the calculations by G{\"u}nther and Wawrzyn (2011) and our
calculations, respectively. The scale in the upper part of
the figure is the logarithm of the Rosseland optical depth
for the models with heating.
}
 \label{test1}
\end{figure}

  To test our code, we computed a model atmosphere
without heating and compared it with
the model obtained with the unmodified ATLAS9
code. The parameters of this model are typical of
CTTS: $T_{eff} =4100$ K, $\log g=4.5,$ solar elemental
abundances. As we see from Fig. \ref{test1} (see the curve
in the lower right corner of the figure), the results
of our calculations based on the two codes for an
atmosphere without external irradiation and $P_0=0$
closely coincide.

%
%
Subsequently, we compared the model computed
using our version of the \texttt{ATLAS9} code with the calculations
performed by G{\"u}nther and Wawrzyn (2011)
using the \texttt{PHOENIX} code, in which hydrogen and
helium are taken into account without assuming
LTE. We considered a star with the same parameters
as those in the preceding case on which blackbody
radiation with $T_{B}=20\,000$ K was incident perpendicular
to its surface, with the ratio of the external
radiation flux $F_{B}$ to the stellar one being 5.31. For
comparison with this model, we set $P_0=0$ in our
code and replaced the shock radiation spectrum by
a blackbody one with the following characteristics:
$$
I_\nu^0(\mu) = F_B\delta(\mu +1)\delta(\varphi), \quad
J_\nu^0 =  \frac{F_B}{4\pi} e^{-\tau_\nu}, \quad
H_\nu^0 = -\frac{F_B}{4\pi} e^{-\tau_\nu},
$$
where $\delta(x)$ is the delta function. In our code, we additionally
specified the microturbulence $v_t=1$ km\,s$^{-1}$
and the parameter describing convection $l/H_p=2.$

As we see from Fig. \ref{test1}, the temperature differences
in the region with $\log P > 2,$ which is of interest in
the problem on heating the atmospheres of CTTS by
shock radiation, do not exceed 250 K. The models
begin to differ greatly at lower values of $P$ to which
a Rosseland mean optical depth $\log\tau_{Ross}<-3.3$ corresponds,
i.e., where the LTE approximation is barely
justified. Therefore, the model computed with the
\texttt{PHOENIX} code must yield more realistic results
than the \texttt{ATLAS9} code. It follows from the aforesaid
that our modified \texttt{ATLAS9} code is quite suitable for
solving the formulated problem.

\section*{Dependence of the vertical hot-spot structure on shock parameters}
As has already been pointed out, when calculating
the heating of CTTS atmospheres, we used
the opacity tables for a microturbulence of 2 km\,s$^{-1}$
and solar elemental abundances. Everywhere below,
unless otherwise specified, we took $\log g=4.0$ and,
when describing convection, $l/H_p=1.25.$ Our calculations
were performed on a mesh of 72 Rosseland
optical depths from $\log \tau_{Ross} = -5.1$ to $\log\tau_{Ross}=2$
with a step $\Delta\log\tau=0.1.$ Thus, the models considered
below differ from one another by the stellar
effective temperatures $T_{eff}$ and shock parameters, i.e.,
by $N_0$ and $V_0.$

Let us first consider how sensitive the structure
of CTTS atmospheres heated by shock radiation is
to the angular dependence $I_\nu^0(\mu).$ For this purpose,
we computed two models describing the heating of a
star with $T_{eff}=4000$ K by radiation from a shock with
$V_0=400$ km\,s$^{-1}$ and $\log N_0=12.0.$ They differ from
each other in that we used dependence (\ref{I-thin}) in the first
case and (\ref{I-const}) in the second case.

\begin{figure}
\begin{center}
\includegraphics[scale=0.5]{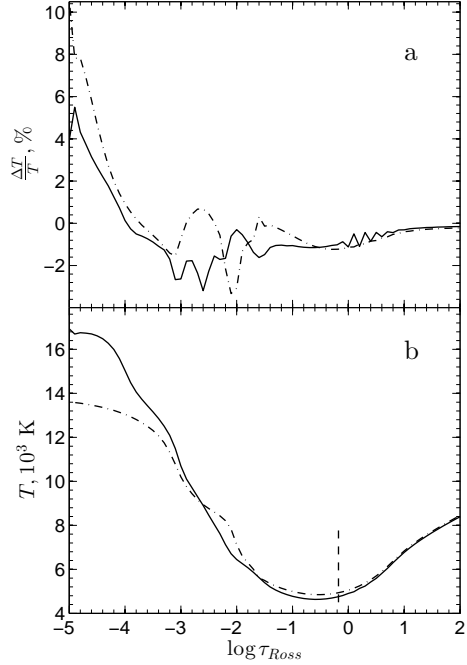}
\end{center}
\caption{(a) Relative temperature difference between two
models in which all parameters are identical but the dependence
$I_\nu^0(\mu)$  is different: it is specified by Eq. (\ref{I-thin}) in one
model and by Eq. (\ref{I-const}) in the other model. The solid line is
for the pair of models with $V_0=400$ km\,s$^{-1}$ and $\lg N_0=12.0;$
the dash-dotted line is for $V_0=200$ km\,s$^{-1}$ and
$\log N_0=12.9.$ The Rosseland mean optical depth is
along the horizontal axis. (b) The temperature distribution
in an atmosphere heated by shock radiation
with an isotropic dependence of the incident radiation (\ref{I-const})
and $V_0=400$ km\,s$^{-1}$ and $\log N_0=12.0$ (solid line),
$V_0=200$ km\,s$^{-1}$ and $\log N_0=12.9$ (dash-dotted line).
The vertical dashed line marks the position of $\tau_{Ross}=2/3$.
}
 \label{test2}
\end{figure}

Since here we calculate LTE spectra, the difference
in temperature distribution for these models is
of chief interest. The solid line in Fig.\,\ref{test2} indicates
the relative temperature difference $\Delta T/T$ of the pair
of models under consideration as a function of the
optical depth $\tau_{Ross}.$ We see that $\Delta T/T\le 3$\,\% in the
region $\log \tau_{Ross} > -4$. An equally
small difference is also obtained when comparing the
pair of models with other identical shock parameters
and $T_{eff}$ of the star but with different laws $I_\nu^0(\mu)$ -- see,
for example, the dash-dotted curve in the same plot
that indicates the result of our calculations for $T_{eff}=4000$ K,
 $V_0=200$ km\,s$^{-1}$, $\log N_0 = 12.9.$ Consequently,
the uncertainty in choosing the law $I_\nu^0(\mu)$ in
the problem on heating of the atmospheres of CTTS
by shock radiation may be considered to be of no
fundamental importance, and we will always assume
below that the shock radiation intensity is isotropic,
i.e., described by law (\ref{I-const}).

Let us now examine how the structure of the atmosphere
being heated changes with shock parameters.
For this purpose, we will fix the star's effective
temperature and consider a pair of models with different
parameters $N_0$ and $V_0$ but with identical accretion
energy fluxes $F_{ac} =\int F_\nu^{sh} {\rm d}\nu \simeq \rho_0 V_0^3/4.$
\footnote{In this case, the radiation flux incident on the stellar atmosphere
will slightly differ for models with different $N_0$ and $V_0$
due to different contributions of the radiation from the preshock
region.}

 The dash-dotted and solid lines in the lower panel
of Fig.\,\ref{test2} indicate the dependence$T=T(\tau_{Ross})$ for the
models with $V_0=200$ km\,s$^{-1}$ , $\log N_0 = 12.9$ and $V_0 =
400$ km\,s$^{-1}$, $\log N_0 = 12.0$, respectively. Both models
were computed for a star with $T_{eff}$ = 4000 K, have the
same $F_ac$, but differ by the spectral composition of
the shock radiation and the pressure $P_0$ at the outer
atmospheric boundary.

We see from Fig.\,\ref{test2} that in the region with $\tau_{Ross}\gtrsim 2/3$
(this optical depth is marked by the vertical line
in the figure), i.e., where the continuum emission
originates, the differences between the models are
comparatively small. In other words, for models
with the same $F_{ac}$, the intensity and spectrum of the
emission continuum should be approximately identical.
At the same time, the differences in structure
become noticeable in the formation region of emission
lines: for a softer shock radiation spectrum, i.e., at
$V_0=200$ km\,s$^{-1}$ in the case under consideration,
the layers with a moderately small optical depth are
heated slightly more strongly that they are for a harder
spectrum, but, on the other hand, the atmosphere for
a soft spectrum at quite $\tau_{Ross}$ is heated to a much
lesser extent.

\begin{figure}
\begin{center}
\includegraphics[scale=0.7]{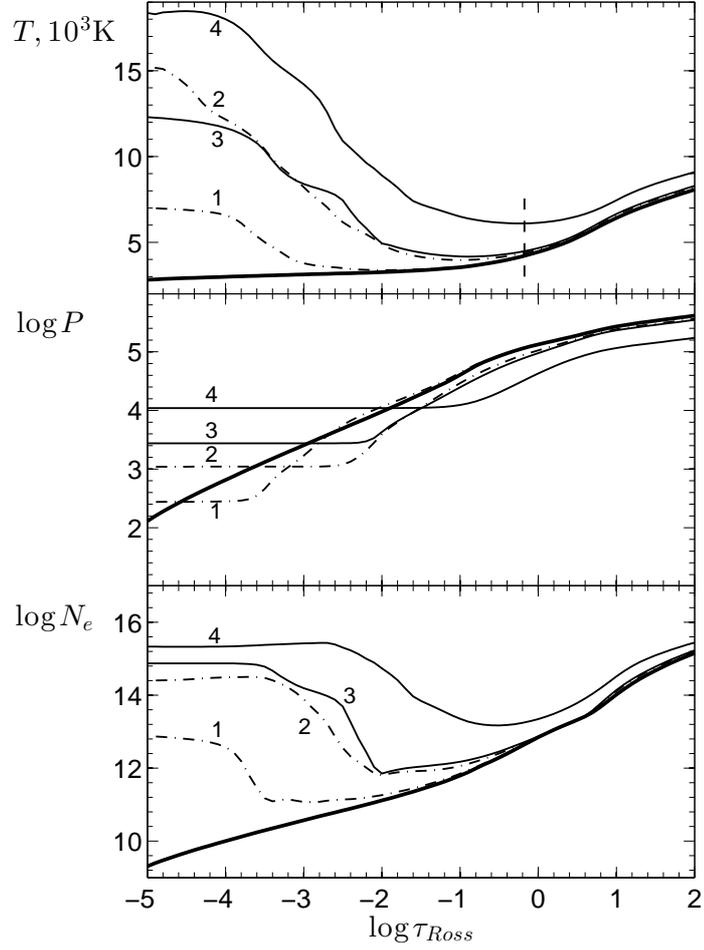}
\end{center}
\caption{
Dependence of the model for the heating of a
stellar atmosphere with $T_{eff}$ = 4000 K and $\log g = 4.0$
on external radiation properties. $\tau_{Ross}$ is the Rosseland
optical depth; the vertical line indicates $\tau_{Ross}=2/3.$
From top to bottom: the temperature ($10^3$ K),
pressure (dyn\,cm$^{-2}),$ and electron density (cm$^{-3})$ distributions.
The thick line indicates the model without
heating, 1 -- V0 = 200 km\,s$^{-1},$ $\log N_0 = 11.5;$ 2 --
$V_0 = 400$ km\,s$^{-1},$ $\log N_0 = 11.5;$ 3 -- $V_0 = 200$ km\,s$^{-1},$
$\log N_0 = 12.5;$ 4 -- $V_0 = 400$ km\,s$^{-1},$ $\log N_0 = 12.5.$
}
 \label{models}
\end{figure}

  It is quite predictable that the structure of the
atmosphere being heated depends mainly on the external
radiation power or, more precisely, on the parameter
$K=F_{ac}/\sigma T^4_{eff}.$
This is illustrated by Fig.\,\ref{models}.
It shows how the gas temperature and pressure and
the electron density change with optical depth $\tau_{Ross}$
for the models that we will use most frequently below.
These models are numbered in the figure in order
of increasing K: (1) $V_0 = 200$ km\,s$^{-1}$, $\log N_0 =
11.5$ $(K \simeq 0.09);$ (2) $V_0 = 400$ km\,s$^{-1}$, $\log N_0 =
11.5$ $(K \simeq 0.76);$ (3) $V_0 = 200$ km\,s$^{-1}$, $\log N_0 =
12.5$ $(K \simeq 0.94);$ (4) $V_0 = 400$ km\,s$^{-1}$, $\log N_0 = 12.5$
$(K \simeq 7.6).$ In all cases, $T_{eff}=4000$ K.

The larger the $K,$ the greater the difference between
the structure of the atmosphere being heated
and the initial one. At the same time, we see from
the figure that the dependence $T=T(\tau_{Ross})$ at $K<1$
in the region with $\tau_{Ross}\gtrsim 2/3,$ i.e., where the continuum
originates, is almost the same as that in an
unperturbed atmosphere. This means that the depth
of photospheric lines can decrease significantly due
to the emission continuum only at $K>1.$  Therefore,
when interpreting the spectra of noticeably veiled
CTTS, Calvet and Gullbring (1998), who disregarded
the emission in lines in their calculations, always
obtained a large $K.$

\subsection*{Computing the Spectrum}
Having determined the structure of the atmosphere
being heated, we computed the spectrum of
the emergent radiation from it using the \texttt{SYNTHE}
code in the \texttt{ATLAS9} package. In contrast to Calvet
and Gullbring (1998), we calculated not the flux $H_\nu,$
but the intensity of the emergent radiation from the
hot spot $I_\nu^{spot}(\mu)$ for 17 values of $\mu$ from 1.0 to 0.01.

In our calculations, we took into account both
the continuum and all the lines of atoms, ions, and
molecules available in the \texttt{ATLAS9} package. The
spectral resolution $R\equiv \lambda/\Delta \lambda$ in
our calculations was $\simeq 600\,000.$

Before reaching the observer, the hot-spot radiation
passes through the material both downstream
and upstream of the shock front. However, it turned
out that in the {\it visible} range at $V_0$ and $N_0$ of interest
to us, the gas falling to the star hardly distorts the
spectrum and intensity of the hot-spot radiation. This
is mainly attributable to a very small optical depth of
the accreting gas in continuum $\lambda > 4000$ \AA\, (Calvet
and Gullbring 1998; Lamzin 1998). As regards the
lines, they must be redshifted in the gas moving toward
the star, while only a very small number of lines
in the visible range have noticeable absorption and/or
emission components in the red wing in the optical
spectra of CTTS (see, e.g., Petrov et al. 2001). The
results of modeling the profiles of such lines are to be
presented in the immediate future.

\begin{figure}
\begin{center}
\includegraphics[scale=0.5]{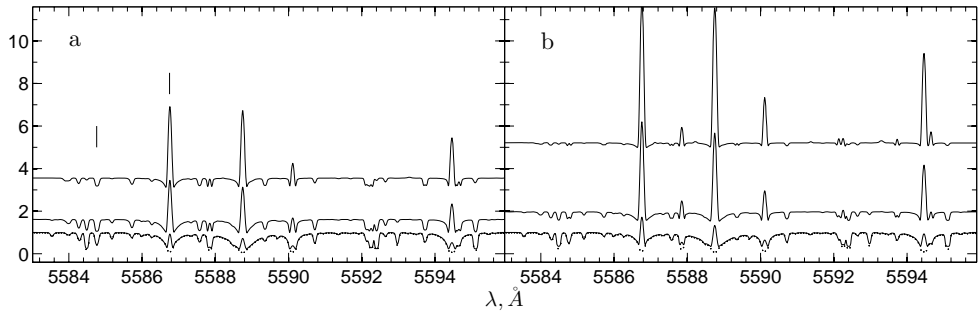}
\end{center}
\caption{
Portion of the spectrum for a plane layer with $T_{eff}=4000$ K (dotted line) and the same layer heated by shock radiation
at different K: for a layer with the normal to its surface making the angles $\alpha=0^o$ (a) and $\alpha=60^o$ (b)
with the line of sight. The three curves in each panel correspond to the spectra of the models with the following parameters
(from bottom to top):
$V_0 = 200$ km\,s$^{-1}$ and $\log N_0 = 11.5$ $(K\simeq 0.09),$
$V_0 = 200$ km\,s$^{-1}$ and $\log N_0 = 12.5$ $(K\simeq 0.76),$
$V_0 = 400$ km\,s$^{-1}$ and $\log N_0 = 12.5$ $(K\simeq 2.6).$
The vertical bars mark the positions of two Fe\,I lines with $gf$ differing by a factor of 160. The layer
radiation intensity $I_\lambda$ normalized to the {\it continuum} intensity of a layer without heating is along the vertical axis.
}
  \label{slab-spectra}
\end{figure}

  Figure \ref{slab-spectra} shows how the radiation spectrum of a
plane-parallel layer with $T_{eff}=4000$ K (dotted curve)
changes as it is heated by shock radiation with $K=0.09,$
0.76, and 2.6 (solid curves). To be more precise,
the dependencies $I_\lambda (\lambda)$ are shown in the left and right
panels of the figure for the cases where the line of sight
makes, respectively, the angles $\alpha=0^o$ and $\alpha=60^o.$
with the normal to the layer. For clarity, the spectra
were normalized to the continuum level $I_c^0$ of a layer
without external heating $(K=0).$

   Let us first consider the behavior of the spectra
for $\alpha=0^o.$ We see that the continuum intensity at
$K=0.09$ (lower solid curve) remained essentially the
same, but the depth of some absorption lines, for
example, Fe\,I $\lambda$ 5586.8 and Ca\,I $\lambda$ 5588.8, decreased
appreciably. The continuum level at $K=0.76$  (middle
solid curve) increased approximately by a factor
of 1.6, while the depth of absorption lines decreased to
an even greater extent and some of them turned into
emission ones. This effect is even more pronounced
in the spectrum at $K=2.6$ (upper curve). If we now
take a look at the right panel of the figure, which
corresponds to the case of $\alpha=60^o,$ then we will see
that the picture did not change qualitatively, but the
continuum level increases faster with increasing $K,$
while the emission lines appear already at $K = 0.09.$

   These peculiarities can be understood by taking
into account the following facts: (1) in the LTE
approximation, the intensity of radiation at a given
wavelength is approximately equal to the value of the
Planck function in a region with an optical depth $\tau_\lambda
\simeq 1;$ (2) a temperature inversion arises in an atmosphere
heated by shock radiation, with the position of the
temperature minimum shifting to increasingly large
$\tau_{Ross}$ as $K$ increases (see the upper panel of Fig. \ref{models}).

  The intensity of the spot radiation in continuum at
$K\simeq 0.1$ is almost equal to $I_c^0,$ because the temperature
$T_{eff}$ at $\tau_{Ross} \simeq 1$ changed very little compared to
what it was in the absence of heating -- see curve 1
in the upper panel of Fig. \ref{models}. The lines are formed at
smaller $\tau_{Ross},$ where the relative rise in temperature
is larger. Therefore, their intensity increases to a
greater extent than it does in the adjacent continuum,
causing the depth of absorption lines to decrease, i.e.,
their veiling. Depending on how large the absorption
coefficient of a line is, it originates in a region with a
temperature that is either lower or higher than that in
the formation region of the adjacent continuum. The
line will appear as an absorption one in the former
case and as an emission one in the latter case. There
are no emission lines at $K=0.1$ in the portion of the
spectrum shown in Fig. \ref{slab-spectra}, but they are present in
other portions of the spectrum -- for example, the
Fe\,II $\lambda$ 5234.6 line, while the emission line also appear
in the chosen portion as $K$ increases. The behavior
of the two Fe\,I lines with $\lambda =5584.8$ and $\lambda = 5586.8,$
for which $\log gf$ is -2.32 and -0.12, respectively, can
serve as an illustration of the aforesaid.

If we look at the layer not along the normal to its
surface but at an angle, then the outermost atmospheric
layers, whose temperature at $\mu=0.5$ is higher
than that at $\mu=1$, will correspond to $\tau_\lambda=1.$
In other words, for the hot spot, just as for the solar chromosphere,
the law $I_\lambda=I_\lambda(\mu)$ corresponds not to limb
darkening but to limb brightening; we see from Fig. \ref{slab-spectra}
that the dependence $I(\mu)$ in the case of lines is steeper
than that for the continuum. As regards the quantitative
differences in continuum level at the same $K$ but
different $\mu,$ it should be remembered that the curves
$I_\lambda(\lambda)$ in the figure are normalized to the continuum
level of an unperturbed atmosphere, where $I_\lambda(0.5) < I_\lambda(1).$

   To compare our calculations with observations, we
assumed here that on the stellar surface there was
only one circular spot within which the shock parameters,
i.e., $V_0$ and $N_0,$ were identical. The position of
the spot in this case is determined only by the angle á
between the normal at the spot center and the line of
sight. Another characteristic of the spot is $f,$ the ratio
of its area to the surface area of the entire star.

The radiation coming to the observer is the sum
of the radiations from the spot and the unperturbed
stellar surface. The observed flux $F_\lambda$ is obtained by
integrating the intensity $I_\lambda(\mu)$ over the solid angle:
$$
F_\lambda = {R_*^2 \over d^2} \,
\int{ I_\lambda(\mu)\, \mu \, \sin \theta \,
{\rm d}\phi \,\, {\rm d} \theta},
$$
where $\theta$ and $\phi$ are the spherical coordinates of the
points on the surface of a star with radius $R_*$ at
distance $d$ from us, and $\mu = \mu(\theta,\phi)$ is the cosine of
the angle $\alpha$ between the local normal to the surface
and the line of sight. In our calculations, we used a
uniform $100 \times 100$ coordinate grid in $\theta$ and $\phi$ for each
cell of which we can write
$$
I_\lambda(\theta,\phi) = \begin{cases}
 I_\lambda^{star}, \text{ if $(\theta,\phi) \notin \Pi$ } \\
\\
 I_\lambda^{spot}, \text{ if $(\theta,\phi) \in \Pi$ }, \\
 \end{cases}
$$
where $\Pi$ is the domain of $\theta$ and $\phi$ in which the hot spot
is specified.


\section*{Comparison of the calculation with observations}

\subsection*{The relative contribution of lines and continuum to the veiling}
\begin{figure}
\begin{center}
\includegraphics[scale=0.7]{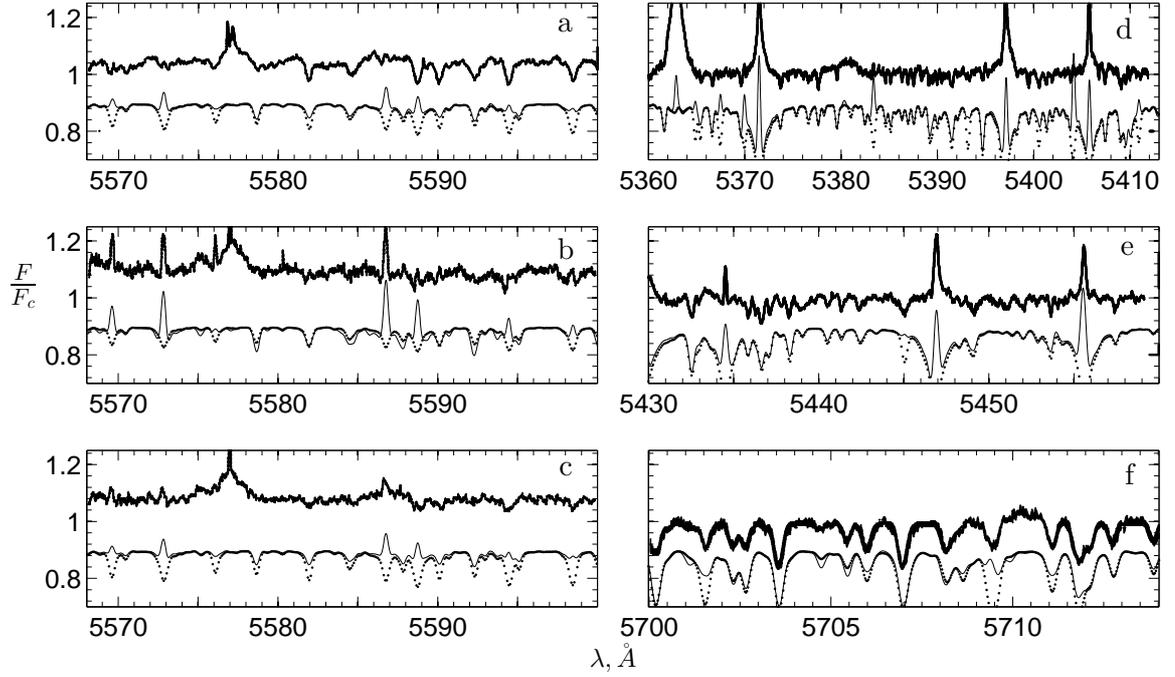}
\end{center}
\caption{
Calculated and observed spectra of heavily veiled CTTS. The thick curve indicates the observed spectrum: (a) RU Lup,
(b) S CrA SE, (c) S CrA NW (Gahm et al. 2008), (d) and (e) DR Tau (Petrov et al. 2011), (f) RU Lup (Stempels and
Piskunov 2003). The thin solid line indicates the spectra of the models from the table, while the dotted line indicates the
spectra of the same models calculated by disregarding the emission in lines. The spectra were normalized to the continuum
level but were shifted vertically for the convenience of comparison.
}
 \label{spectra-compar}
\end{figure}

Evidence for the presence of emission lines inside
photospheric absorption lines has been found in stars
with heavily veiled spectra: RU Lup, S CrA NW,
S CrA SE (Gahm et al. 2008), DR Tau (Petrov
et al. 2011). For each of these stars, we were able
to select a model whose spectrum, as can be seen
from Fig. \ref{spectra-compar}, is rather similar to the observed one, at
least in that the lines exhibiting an emission feature
in the observed spectra also exhibit emission features
in the models\footnote{ Except for the [O\,I] 5577 line originating in the CTTS wind;
therefore, it must have no analogue in the calculated spectra.}.
The observed spectrum is shown in
the upper part of each panel in the figure, and the thin
solid line below indicates the model spectrum for a
star with a hot spot whose parameters are given in the
table. Both spectra were normalized to the continuum
level, but, for clarity, they were shifted from $F_\lambda/F^c_\lambda=1$
vertically. In addition, the calculated spectra were
broadened by its convolution with a Gaussian with
$\sigma=7.5$ km\,s$^{-1}$ in order that their spectral resolution
be the same as that of the observed ones.

%
%


\begin{table}
  \caption{Parameters of the models for heavily veiled stars}
 \label{highveiled}
\begin{center}
\begin{tabular}{|c|l|c|c|c|c|c|c|c|}
\hline
\# & Star & $r$ & $f$ & $\log N_0$, cm$^{-3}$ & $V_0,$ km\,s$^{-1}$ & $T_{2/3}$, K &
$R_*/R_\odot$ & $\dot{M},$ $M_{\odot}$ yr$^{-1}$ \\
\hline
1 & RU Lup$^a$   & 6  & 0.12 & 12.5 & 400 & 6100 & 1.7$^d$  &$9\times10^{-8}$\\
2 & S CrA NW$^a$ & 8  & 0.12 & 12.5 & 400 & 6100 & 1.2$^e$  &$5\times10^{-8}$\\
3 & S CrA SE$^a$ & 5  & 0.15 & 12.5 & 300 & 5200 & 1.5$^e$  &$7\times10^{-8}$\\
4 & DR Tau$^b$   & 2.5& 0.10 & 12.5 & 300 & 5200 & 1.5$^b$  &$4\times10^{-8}$\\
5 & RU Lup$^c$   & 2.2& 0.15 & 12   & 400 & 4800 & 1.7$^d$  &$4\times10^{-8}$\\
\hline
 \multicolumn{9}{p{16cm}}
{\footnotesize {\bf Note.} $T_{2/3}$ is the hot-spot temperature at $\tau_{Ross}=2/3,$ $R_*$ is the stellar radius, $N_0$ and $V_0$ are the pre-shock gas density and velocity, $r$ is the average veiling level in the range of the corresponding spectrum (Fig. \ref{spectra-compar}), $\dot{M}$ is the accretion rate. The spectra and parameters of the stars were taken from the following papers: (a) Gahm et al. (2008); (b) Petrov et al. (2011); (c) Stempels and Piskunov (2003); (d) Stempels and Piskunov (2002); (e) Carmona et al. (2007).
}
\end{tabular}
\end{center}
\end{table}
%

As we see from the figure, the weak emission
lines of the atmosphere being heated to a certain
extent blend the photospheric absorption lines, while
the strongest ones manifest themselves in the spectra
as the so-called narrow emission components
(see the Introduction). It is important to note that
the calculated intensity of the narrow components
in metal lines is much lower than the observed intensity
of the corresponding emission lines. This
is entirely consistent with the conclusion that the
emission lines of metals consist mainly of the so-called
broad component forming outside the hot spot:
Batagla et al. (1996) and Petrov et al. (2001) reached
this conclusion by analyzing the profile variability,
while Dodin et al. (2012) showed that for RW Aur
the magnetic field in the formation region of metal
emission lines is much weaker than that in the hot
spot.

   Below, we will discuss how well the model parameters
from the table can describe the corresponding
stars, while now we will ascertain to what extent
and under what conditions allowance for the emission
lines influences the veiling. The first impression can
be gained even from Fig.\,\ref{spectra-compar}, in each panel of which
the dotted curve indicates the spectrum of the corresponding
model from the table computed using the
\texttt{SYNTHE} code but without any allowance for the
spectral lines. In other words, this is the spectrum
of a star with a hot spot veiled only by the emission
continuum.

The value of the following quantities averaged over
the spectral range of interest to us is traditionally used
as a measure of veiling of the CTTS spectra:
$$
r(\lambda) = {EW_0(\lambda) \over EW(\lambda)} - 1.
$$
They show the extent to which the equivalent widths
of photospheric lines $EW$ in the spectrum of CTTS
differ from the equivalent widths $EW_0$ of the same
lines in the spectrum of a comparison star. Up until
now, the comparison star is chosen in such a way
that the sum of its spectrum and a constant-intensity
"veiling continuum"{} in a comparatively narrow wavelength
range fits best the spectrum of the star being
studied (see, e.g., Petrov et al. 2001). The difference
in $r$ for different lines was assumed to be due to
observational errors and/or not quite an appropriate
choice of the comparison star.

For our models, $r$ were determined by comparing
two computed spectra: stars with and without
a hot spot. In this case, we took two overlapping
spectral lines as one if there were no local maxima
inside the blend profile (the lines were indistinguishable);
otherwise we considered the lines separately
and regarded the point of maximum as the boundary
between them. Recall that we compute the models in
the LTE approximation, which definitely breaks down
at very small $\tau_{Ross},$ i.e., in the formation region of the
strongest lines. Therefore, in determining $r$ we used
only sufficiently weak lines originating in layers where
the LTE hypothesis seems justified. On the other
hand, there was no point in taking into account very
weak lines $(EW<0.03$ \AA\,), because they are masked
by noise in the observed spectra.

\begin{figure}
\begin{center}
\includegraphics[scale=0.7]{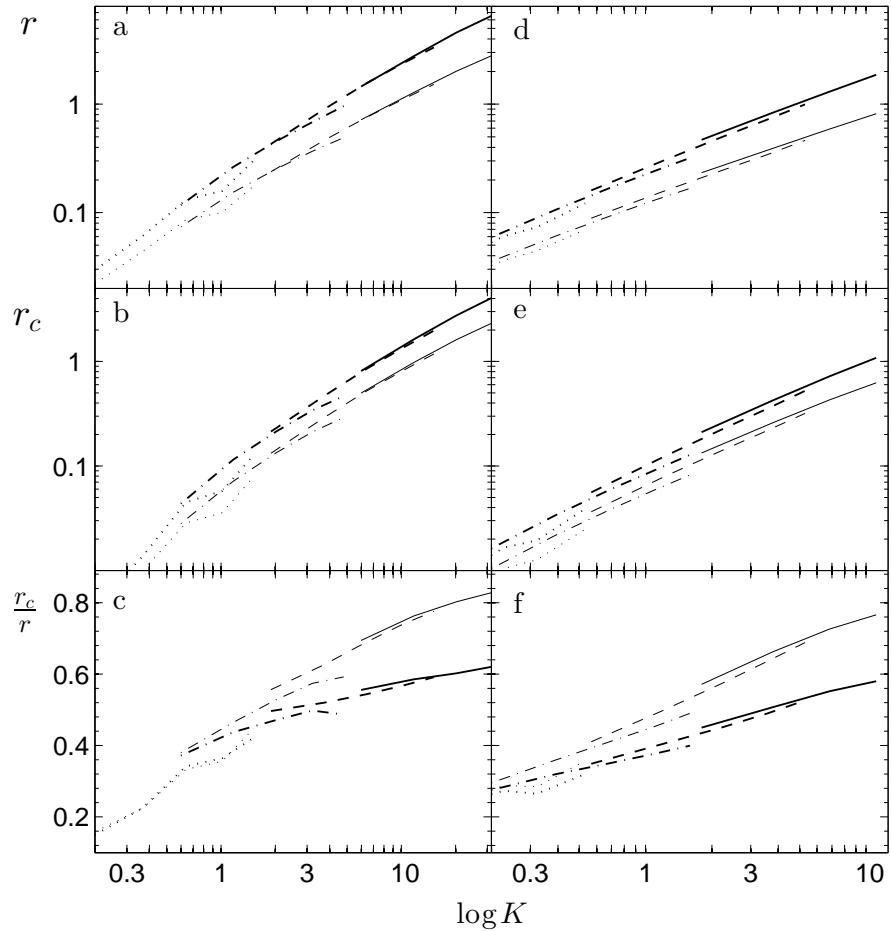}
\end{center}
\caption{
Quantities $r$ (upper row), $r_c$ (middle row), and their ratio (lower row) in the range 5500-6000 \AA\, versus relative shock
radiation power in the case of a spot that occupies 6\% of the surface of a star with $T_{eff}=4000$ K (left column) and $T_{eff}=5000$ K
(right column). Each line corresponds to a series of models with identical $N_0$ but different $V_0$ from the range 200-400 km\,$^{-1}.$
The solid, dashed, dash-dotted, and dotted lines are for $\log N_0 = 13.0,$ $12.5,$ $12.0,$ and $11.5,$ respectively. The thicker and
thinner lines correspond to a spot observed at $\alpha=0^o$ and $\alpha=60^o,$ respectively.
}
 \label{r-rc-figure}
\end{figure}

  As the external radiation power increases, the degree
of veiling of the spectrum for a star with a hot
spot grows. Figure \ref{r-rc-figure} gives an idea of the quantitative
magnitude of the effect. As an example, it shows the
dependence $r(K)$ in the range 5500-6000 \AA\, for a
moderate-size hot spot $(f=6\,\%)$ on the surface of a
star with $T_{eff}=4000$ K (Fig. \ref{r-rc-figure}a) and $T_{eff}=5000$ K
(Fig. \ref{r-rc-figure}d). The thick and thin lines correspond to the
cases where the spot symmetry axis makes, respectively,
the angles $\alpha=0^o$ and $\alpha=60^o$ with the line of
sight.

%
%

   If the equivalent widths changed only through a
change in the continuum level, then the veiling could
be characterized by the parameter
$$
r_c(\lambda) = { F_\lambda \over F^0_\lambda } - 1,
$$
where $F_\lambda$ and $F^0_\lambda$ are the continuum fluxes for the
accreting and nonaccreting stars, respectively. The
ratio $r_c/r$ shows what fraction of the "true" veiling is
attributable to the emission continuum alone.

Figures \ref{r-rc-figure}b, \ref{r-rc-figure}e and
\ref{r-rc-figure}c, \ref{r-rc-figure}d show how $r_c$ and $r_c/r,$
respectively, change as the shock radiation power
increases for the same spots and in the same spectral
range. We see that both $r_c$ and $r_c/r$ increase with $K.$
This means that the contribution of lines to the spectrum
veiling is largest when r is comparatively small
(in our case, $r_c/r < 0.5$ at $r<0.3-0.4).$ However,
we have already seen this for the radiation of a plane
layer as an example (see the previous section and
Fig. \ref{slab-spectra}).

  In the opinion of Petrov et al. (2011), if the contribution
of lines to the veiling is disregarded, then
the accretion rate will be overestimated most for stars
with large $r$. It follows from Fig.\,\ref{r-rc-figure} that, in general,
this is not true. Suppose, for example, that in the
case of a star with $T_{eff}=4000$ K and a spot with
$f=0.06,$ $r = 0.1$ was obtained from observations at
$\alpha=0^o,$ which, neglecting the contribution of lines, is
mistaken for $r_c.$ Using the figure, we find that as a
result of the misinterpretation, we will obtain $K$ that
is twice the correct one. However, if the observed
degree of veiling for the same star is $r = 3,$ then the
error in $K$ will be not larger but, on the contrary,
smaller approximately by a factor of 1.5. Assuming,
for simplicity, that $V_0$ was correctly determined from
observations, we will obtain the same errors for the
accretion rate as well.

   It may appear that the lines for stars with a high
veiling level should be considered as an insignificant
correction. However, this is not the case, and we will
now discuss a number of qualitatively new effects that
arise when the emission in lines is taken into account.

\begin{figure}
\begin{center}
\includegraphics[scale=0.7]{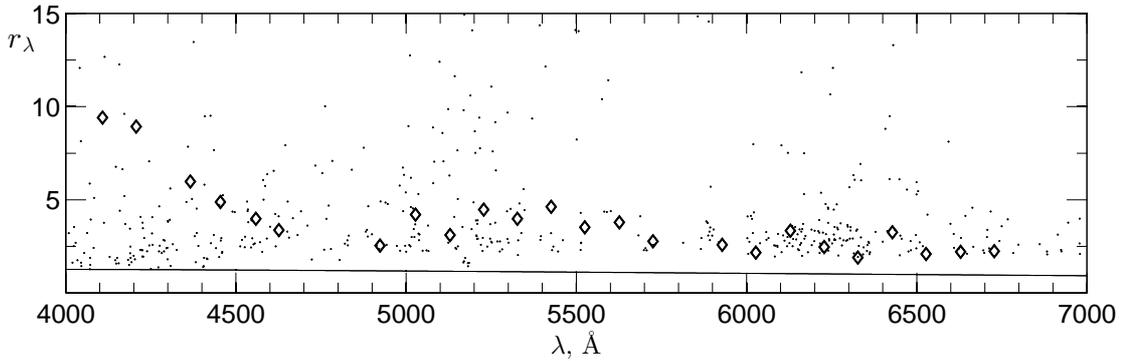}
\end{center}
\caption{
Wavelength dependence of the veiling. For model no. 5 from the table, the dots indicate the values of $r$ obtained from
individual lines, while the solid line indicates the dependence $r_c(\lambda).$ The diamonds represent the average veiling in narrow
regions of the spectrum for RU Lup determined by Stempels and Piskunov (2003).
}
 \label{rulup15}
\end{figure}

Figure \ref{rulup15} shows how $r$ and $r_c$ change in different
regions of the spectrum for model no. 5 from the
table. We chose this model for comparison with
the spectrum of RU Lup taken from Stempels and
Piskunov (2003): a small region of this spectrum is
presented in Fig. \ref{spectra-compar}f. In accordance with the aforesaid,
all points in the figure lie above the line $r_c = r_c(\lambda);$
in this case, $r$ averaged over small regions of the
spectrum are considerably larger than $r_c,$ i.e., there
is veiling mainly by lines, not by the emission continuum,
especially in the short-wavelength part of the
spectrum.

%
%

 As we see from the figure, the values of $r$ can
differ by several times even for closely spaced lines.
The quantity $r$ has a particularly large scatter in the
wavelength range 5000-5500 {\AA}. As a result, its mean
value in this range increases, with such a peculiarity
taking place in all of our computed models. Thus,
the local maximum in the dependence $r=r(\lambda)$ near
$\lambda \approx 5300$ {\AA} found by Stempels and Piskunov (2003)
for RU Lup and by Hartigan et al. (1989) for BP Tau
cannot be considered as evidence for a nonmonotonic
energy distribution of the veiling continuum in the
visible spectral range.

  Unless we take into account the fact that the veiling
by lines changes the depth of photospheric lines to
a very different degree even within a narrow spectral
range, the universally accepted method of determining
the effective temperatures of CTTS (Hartigan
et al. 1989) by comparing $EW$ of absorption lines in
the spectrum of the program and comparison stars
must yield erroneous results. Using this technique,
we processed our computed model spectra for a star
with a hot spot at two spot positions relative to the
observer: $\alpha=0^o$ and $\alpha=60^o.$ The error of the
method by Hartigan et al. (1989) can be estimated
by comparing $T_{eff}$ obtained in this case with the value
taken when computing the model. We used our
computed grid of spectra for stars without a spot with
$\log g=4.0$ and $T_{eff}$ from 3500 to 5750 K with a step
of 250 K as comparison spectra.

 The results turned out to be ambiguous: the difference
$\Delta T_{eff}$ between the calculated and actual values
of $T_{eff}$ most commonly did not exceed 250 K, i.e.,
no more than one spectral subtype, but, in some
cases, the differences were considerably larger. For
example, for model no. 5 from the table at $\alpha=0^o,$
we found $\Delta T_{eff} = +1750$ K over the spectral region
4500-5000 {\AA}, $\Delta T_{eff} = -250$ K in the range 5000-5500 {\AA},
and $\Delta T_{eff} =0$ K over regions 500 {\AA} in width
at $\lambda >5500.$ In models with a moderately large
degree of veiling of the stellar spectrum, the error in
$T_{eff},$ as would be expected, is small, but for models
with an appreciable veiling, $\Delta T_{eff}$ changes from model
to model without any apparent regularity. This is
probably because as the degree of veiling increases,
some of the photospheric lines either become very
weak or turn into emission ones and we cease to
take them into account when calculating $r.$ In other
words, in different models, $r$ within the same spectral
range, in general, is calculated from different set of
lines. Therefore, it is reasonable to assume that for
all CTTS with an appreciably veiled spectrum (see,
e.g., Gullbring et al. 1998), $T_{eff}$ and the spectral type
were determined with an error whose value is difficult
to predict, especially since the observed spectra of
CTTS, in contrast to the model ones, are distorted by
noise.

   The error in determining $T_{eff}$ for CTTS with an
appreciably veiled spectrum leads to an error in
estimating the interstellar extinction toward these
stars. Therefore, note that $A_V$ determined from optical
CTTS spectra systematically exceed $A_V$ found
by analyzing ultraviolet spectra (see Lamzin (2006)
and references therein). Lamzin (2006) provided
arguments that some systematic error built in the
technique of determining $A_V$ from optical spectra
is responsible for the effect. It is reasonable to
assume that this error results from the neglect of the
contribution of lines to the veiling.

\begin{figure}
\begin{center}
\includegraphics[scale=0.7]{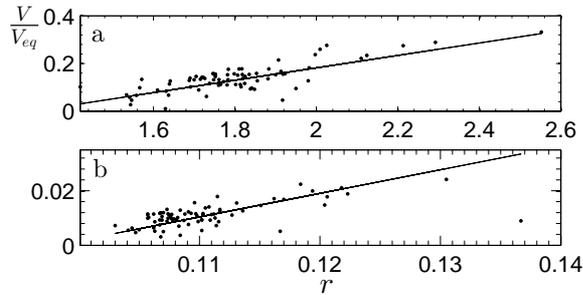}
\end{center}
\caption{
Radial velocity of photospheric lines versus veiling
in them: (a) for model no. 3 from the table $(f = 0.15)$ at
$\alpha=60^o,$ the dots represent individual lines, the straight
line is an analytical fit: $V/V_{eq} = 0.3\times(r-1.3)\sin \alpha;$
(b) the same for a spot with $f = 0.01$, a fit:
$V/V_{eq} = 1.0\times(r-0.1)\sin \alpha .$
}
 \label{vradial}
\end{figure}

   As the star rotates, the radial velocity of the hot-spot
lines changes, causing the centroid of the veiled
photospheric lines to be shifted periodically. This is
perceived as variability of their radial velocity $V_r$ (Zaitseva
et al. 1990; Petrov et al. 2001, 2011). However,
when this effect is interpreted quantitatively, it should
be remembered that all photospheric lines are veiled
to a different degree and, hence, the $V_r$ variability
amplitude will depend on the set of lines used. This is
illustrated by Fig. \ref{vradial}, in which $V_r$ of photospheric lines
from the range 5500-6000{\AA} normalized to the star's
equatorial rotation velocity $V_{eq}$ is plotted against their
degree of veiling $r.$

  The upper panel in the figure presents the results of
our calculations for model no. 3 from the table in the
case where a large spot $(f = 0.15)$ is located at the
equator and is observed at $\alpha = 60^o,$ while the lower
panel presents the results for the same parameters of
a spot but with a smaller size $(f = 0.01).$ The dependence
$V_r(r)$ is well fitted by a straight line whose
equation can be easily derived analytically. This requires
finding the centroid of two Gaussians that
approximately reproduce the absorption and emission
parts of the line, which gives $V_r = (r-r_c) V_s\,\xi,$ where
$r$ is the total veiling, $r_c$ is the continuum veiling,
$V_s=V_{eq}\sin{i}\sin{\theta}\sin{\phi}$ is the line-of-sight projection
of the velocity of the surface element with a spot, $\xi$
is a factor approximately equal to one for a compact
accretion zone $(f\lesssim 0.01)$ and 0.3 for $f=0.15$.

%
%

The presence of emission components inside absorption
lines also affects the estimate of the star's
projected equatorial rotation velocity $v\sin{i}.$ The full
width at half maximum of the line $(FWHM)$ increases
when the emission component is at the line
center (the hot spot passes through the central meridian),
but the photospheric line becomes narrower
when the emission component is in one of the wings.
The variability of $v\sin{i}$ in the spectrum of RW Aur
(Petrov et al. 2001) is probably attributable precisely
to this effect.

\begin{figure}
\begin{center}
\includegraphics[scale=0.7]{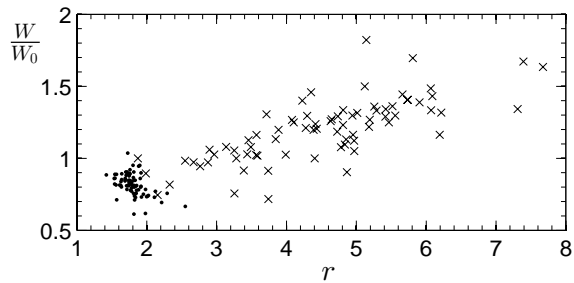}
\end{center}
\caption{
Line broadening versus veiling for model no. 3
from the table.  $W/W_0$ is the ratio of $FWHM$ for a
line in the spectrum of an accreting star to its value in
the spectrum of a star without a spot, $r$ is the veiling.
The equatorial rotation velocity is $V_{eq} = 12$ km\,s$^{-1}.$
The crosses and dots indicate a hot spot observed at $\alpha=0^o$
and $\alpha=60^o,$ respectively.
}
  \label{vsini}
\end{figure}

Just as for the radial velocities, the result of measuring
$v\sin{i}$ depends on which photospheric lines
will be chosen. Figure \ref{vsini} shows how the ratio of
$FWHM$ for a line in an accreting star to $FWHM$
for the same line in a star without a spot changes
with $r.$ The results are presented for model no. 3 from
the table at $\alpha=0^o$ and $\alpha=60^o,$ while the equatorial
rotation velocity was taken to be 12 km\,s$^{-1}.$ For simplicity,
we assumed the spot center to lie at the stellar
equator and the inclination of the rotation axis to the
line of sight to be $i=90^o$. The lines were chosen
from the range 5500-6000 {\AA} in such a way that the
emission part did not distort the profile too strongly
and it could be taken as an ordinary photospheric line
in the observed spectrum.

%
%

  We see from the figure that depending on the
choice of lines from which $v\sin{i}$ is determined, the
values obtained for this quantity can differ several fold.
This suggests that all of the published values
of $v\sin{i}$ for CTTS with a noticeably veiled spectrum
have a systematic error whose value is difficult to
estimate. Precisely this is probably responsible for
the problems that arise in attempting to reconcile
the rotation period of the star, its radius, and $i$ and
$v\sin{i}$ between themselves for DI Cep and RW Aur
(Gameiro et al. 2006; Dodin et al. 2012).

%
\subsection*{Reliability of determining the hot-spot parameters}

  Here, we consider a comparatively simple model --
a circular homogeneous spot. However, the calculations
by Romanova et al. (2004) show that even in
the simplest case of a dipole stellar magnetic field, the
cross section of the accretion stream has a complex
shape, while the distributions of $N_0$ and $V_0$ in this
cross section are very nonuniform. Nevertheless, even
our simple model allows us to see how large the
errors in the parameters characterizing the accretion
onto CTTS ($\dot M,$ $f,$ etc.) are if we use inadequate quality
spectroscopic observations and disregard the
emission in lines when interpreting the veiling.

  Let us first show that several models that reproduce
the observed spectra with approximately the
same accuracy can be selected within the approximation
used. One of the main factors leading to an
ambiguity in determining the accretion parameters is
the spot position on the stellar surface at the time
of observation. Let us illustrate this assertion using
the portion of the spectrum for RU Lup displayed in
Fig. \ref{spectra-compar}a as an example. The observed spectrum of
the star taken from Gahm et al. (2008) is compared
in this panel with the model corresponding to a spot
with $\alpha=0$ and $f=0.12.$ The remaining model parameters
are given in the first row of the table.

\begin{figure}
\begin{center}
\includegraphics[scale=0.65]{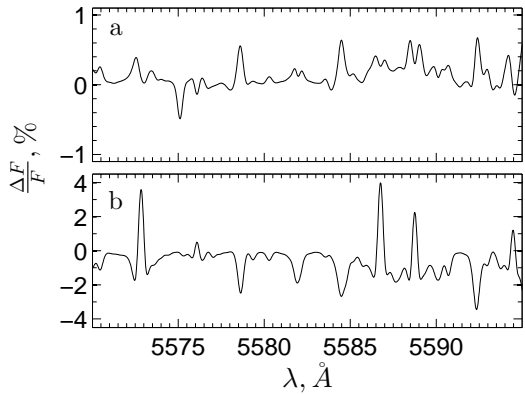}
\end{center}
\caption{
Difference between the spectra $F_\lambda(\lambda)$ for two
models with different parameters. (a) For models in which
only $f$ and $\alpha$ differ. These parameters are, respectively,
0.120 and $0^o$ for the first model and 0.245 and $60^o$ for
the second model. The spectra were broadened by their
convolution with a Gaussian to correspond to the observed
spectrum of RU Lup shown in Fig.\,\ref{spectra-compar}.
(b) For models in which only $V_0$ and $\lg N_0$ differ at the same $K$.
These parameters are $400$ km\,s$^{-1}$ and $12.0$ for the first
model and $200$ km\,s$^{-1}$ and $12.9$ for the second model.
The spectra were broadened by their convolution with a
Gaussian at $\sigma=7.5$ km\,s$^{-1}.$
}
 \label{spectr1}
\end{figure}

Figure \ref{spectr1}a shows the relative difference between
the spectra for this model and the model in which $f=0.245$ and
$\alpha = 60^o,$ while the remaining parameters
are the same. In these models, the accretion rates
$\dot M$ differ approximately by a factor of 2, but we see
from the figure that their spectra coincide to within
1\,\% in the presented portion. To choose between
the two models, observational data with a resolution
$R>40\,000$ and a signal-to-noise ratio $(S/N)$  of at
least 200 should be available. This shows that the
published accretion rates estimated from only one
spectrum and, what is more, of comparatively low
quality cannot be considered reliable even if we forget
that the contribution of lines to the veiling was
disregarded in this case.

  Note that the position of the emission peak inside
an absorption line, which must change periodically
due to the star's rotation about its axis, can serve as
additional information that allows the angle $\alpha$ to be
determined. Such information can be extracted only if
several spectra taken at different positions of the spot
relative to the observer are available.

%
%

   Since the structure of the atmosphere being
heated is determined mainly by the incident radiation
power, the spectra of models with identical $K$ are very
similar. This is illustrated by Fig. \ref{spectr1}b. It shows the
relative difference between the spectra of two models
with identical $K$ but different $V_0$ and $\log N_0,$ which
are, respectively, $400$ km\,s$^{-1},$ $12.0$ and $200$ km\,s$^{-1},$
$12.9.$\footnote{The remaining parameters in these models are also identical:
 $T_{eff}=4000$ K, $\log g=4.0,$ $f=0.1,$ $\alpha=0.$}
The accretion rates in these models differ by
a factor of 4, but the relative difference between $F_\lambda$ in
the spectral range under consideration is $<4$\,\%. Such
a difference can be seen only at a spectral resolution
$R>30\,000$ and $S/N>50.$

  It also follows from our calculations that the spectra
of models with the same $\dot{M}$  but different $f$ and $N_0$
at a given $V_0$ can be distinguished under approximately
the same requirements for the quality of spectroscopic observations
as those when comparing the models with the same $K.$

  For our comparison with the spectra of CTTS in
Fig. \ref{spectra-compar}, we chose models whose spectra were qualitatively
similar to the observed ones: within the model
of a homogeneous circular spot, there is no point in
achieving the best quantitative agreement between
the calculated and observed spectra. In all cases, we
took $T_{eff}=4000$ K and $\log g=4.0$ Since the models
with greatly differing parameters can have very similar
spectra, the quantities given in the table cannot be regarded
as the true characteristic of the corresponding
CTTS. Using the published radii of these stars, we
estimated the accretion rate for them. Although these
estimates are exclusively illustrative, it is worth noting
that compared to $\dot M$ obtained without allowance
for the veiling by lines (Gullbring et al. 2000; Lamzin
et al. 1996), our values are lower by a factor of 3-10.

  Thus, even to determine the parameters of a circular
homogeneous spot $(N_0,$ $V_0,$ $f,$ $\dot M)$ and its position
on the stellar surface, several high-quality spectra
that must be taken during the star's complete turn
about its axis should be available. The effective temperature
of the star, which is not known in advance for
CTTS, should also be determined simultaneously. In
fact, however, an even more complex problem should
be solved: to determine the shape of the accretion spot
and the distribution of $N_0$ and $V_0$ inside it. Basically,
we are talking about the Doppler mapping of CTTS
that should be based on model atmospheres heated
by shock radiation. Numerous photospheric lines of
metals whose depth changes differently due to the
emission components even within a narrow spectral
range should be used in such a mapping.

   Our calculations are the first fundamentally important
step in solving this problem: we not only have
taken into account the hot-spot emission in lines for
the first time but also, in contrast to Calvet and Gullbring
(1998), calculated not the flux but the intensity
of the spot radiation in different directions, without
which no Doppler mapping is possible in principle.
The next step is to take into account the departures
from LTE when calculating the intensity of the lines
that originate in a region with $\log \tau_{Ross}>-3,$ where
there is no need to take into account the non-LTE
effects when calculating the atmospheric structure,
i.e., the dependencies of the ionization fraction and
$T,$ $N_e$ on $\tau_{Ross}.$

  Allowance for the departures from LTE for these
lines may turn out to be important in interpreting
the spectra: in particular, P.P. Petrov (private communication)
found that the photospheric Ca\,I lines in
the spectrum of RW Aur A, in contrast to the lines
of other metals, are veiled only by the continuum.
Our LTE calculations do not allow this effect to be
explained. However, it can be related to a calcium
deficiency in the accreting gas. Indeed, as can be
seen, for example, from Fig. 5 in the review by Spitzer
and Jenkins (1975), the calcium deficiency can be
fairly large in the interstellar medium.


\section*{Conclusions}

   Petrov et al. (2001), Gahm et al. (2008), and
Petrov et al. (2011) provided arguments that the decrease
in the depth of photospheric lines in the spectra
of CTTS was due to not only the presence of an
emission continuum but also a partial filling of absorption
lines with emission ones. As has been shown
here for the first time, this effect stems from the fact
that a stellar atmosphere heated by shock radiation
at the accretion column base (the so-called hot spot)
radiates in both continuum and lines, because the
temperature above the spot photosphere increases
outward. It follows from our calculations that the
strongest of these lines manifest themselves in the
spectra of CTTS as the so-called narrow emission
components, while the weaker ones to a certain extent
blend the photospheric lines.

   The models of stellar atmospheres heated by external
radiation were also computed before us, but
the results of these computations could not be used
to determine the radiation spectrum of the hot spots
on CTTS for the following reasons. In the works
devoted to the reflection effect in binary systems, the
external radiation spectrum differs greatly from the
radiation spectrum of the accretion shock, while the
external radiation source is far from the stellar surface.
Therefore, this radiation is appreciably diluted and
the pressure at the outer boundary of the irradiated
atmosphere is zero. Calvet and Gullbring (1998)
calculated a series of model atmospheres heated by
accretion shock radiation with a nonzero pressure at
the outer boundary. However, they disregarded the
emission in lines when calculating the spectrum of
these atmospheres, because they assumed the veiling
in CTTS to be attributable to the continuum alone.

   Our calculations of the structure and spectrum
of the hot spot were performed using the \texttt{ATLAS9}
code (Kurucz 1970) that we modified in the LTE approximation
for a plane-parallel layer with solar elemental
abundances. Test calculations showed that
the code worked properly. The main parameters of
the problem turned out to be the velocity $V_0$ and
density $N_0$ of the pre-shock accreting gas and the
stellar temperature $T_{eff}.$ The spot spectrum depends
mainly on the ratio $K$ of the external radiation power
to the radiation power of an unperturbed atmosphere;
the spot emission occurs mainly in lines at small $K,$
while a noticeable emission in continuum appears at
larger $K.$ At identical $K$ but different $V_0$ and $N_0,$
the differences between the spot spectra are small.
However, it should be remembered that the shape of
the spectra in the LTE approximation is determined
primarily by the dependence $T=T(\tau_{Ross}).$ Therefore,
takin into account the departures from LTE can
change significantly this conclusion.

   Assuming that there is one circular spot within
which $V_0$ and $N_0$ are the same on the surface of a star
with $T_{eff}$ from 3750 to 5000 K, we calculated how the
resulting spectra of the star$+$spot should appear at
different relative sizes of the spot $f$ and its positions
relative to the observer characterized by the angle $\alpha$
between the line of sight and the spot symmetry axis.
For each of the stars in which the veiling by lines was
detected (Gahm et al. 2008; Petrov et al. 2011), we
were able to select a model with a spectrum similar
to the observed one, at least in that the lines exhibiting
an emission feature in the observed spectra also
exhibit emission features in the models.

   Petrov et al. (2011) pointed out that the accretion
rate for CTTS with a heavily veiled spectrum
could be overestimated by neglecting the contribution
of emission lines to the veiling. Our calculations
show that the accretion rate can be overestimated
by several times, even for stars with a comparatively
weakly veiled spectrum. It turned out that as a
result of the veiling by lines, models with distinctly
different parameters could have very similar spectra.
Therefore, even in the case of a circular homogeneous
spot, to determine $N_0,$ $V_0,$ $f$, and the spot position
on the stellar surface, it is necessary to have several
high-quality spectra that should be taken during the
star's complete turn around its axis. Concurrently,
the effective temperature of the star should also be
determined, whose errors without allowance for the
veiling by lines can be rather large.

   The spot motion relative to the observer as the star
rotates around its axis causes the positions of emission
components inside absorption lines to be shifted.
This is perceived as variability of the star's radial
velocity. Concurrently, the width of photospheric lines
also changes, which appears as variability of $v\sin i.$ It
is important to emphasize that the magnitude of the
effect depends on precisely which photospheric lines
will be chosen for measurements, because the degree
of line veiling $r$ can differ by several times even within
a comparatively narrow spectral range.

   It follows from our calculations that the scatter
of $r$ in the visible range is particularly large in the
wavelength range 5000-5500 {\AA}, as a result of which
its mean value increases in this range. So far the
contribution of lines to the veiling has been disregarded;
the presence of a local maximum in the dependence
$r=r(\lambda)$ near $\lambda \approx 5300$ {\AA} was interpreted
as a nonmonotonic spectral energy distribution of the
emission continuum whose cause was unclear.

     Without allowance for the emission lines, $r$ was a
useful characteristic of the relative hot-spot radiation
power. However, if the lines are taken into account,
then the meaning of $r$ turns out to be by no means
obvious. This means that when the spectra of CTTS
are described, it makes sense to specify the degree
of veiling only by additionally specifying the set of
photospheric lines from which $r$ was determined.

   In reality, the hot spot is definitely noncircular
in shape and the distribution of $N_0$ and $V_0$ in the
cross section of the accretion stream is nonuniform.
To reconstruct the actual picture of accretion onto
CTTS, these stars should be Doppler mapped based
on model atmospheres heated by shock radiation.
Numerous photospheric lines of metals whose depths
change differently due to the emission components
should be used in mapping.

   Our calculations are the first step in solving this
problem: we not only have taken into account the
hot-spot emission in lines for the first time but also,
in contrast to Calvet and Gullbring (1998), have calculated
not the flux but the intensity of the spot radiation
in different directions, without which no Doppler
mapping is possible in principle. In the immediate
future, we are planning to clarify the role of departures
from LTE for the lines that veil the photospheric lines,
but not so strongly as to turn them from absorption
lines into emission ones.

  As regards the helium and metal lines that are in
emission in the spectra of CTTS, as our calculations
confirm, they consist mainly of the so-called broad
component forming outside the hot spot. Subtracting
the narrow components associated with the spot from
the observed emission line profiles, we hope to obtain
information about the profiles of the broad components
that will make it possible to clarify the velocity
field and physical conditions in their formation region.

\bigskip

ACKNOWLEDGMENTS. We wish to thank L.I. Mashonkina and P.P. Petrov
for useful discussions. This work, just as the previous
one (Dodin et al. 2012), was supported by the
Program for Support of Leading Scientific Schools
(NSh-5440.2012.2).

\section*{Appendix. Calculating the structure of the pre-shock zone}

   Lamzin (1998) calculated the shock radiation in
the range of wavelengths shorter than $\simeq 2000$ {\AA},
where the radiation intensity from the photospheres
of CTTS is low. In the optical range that we consider
here, apart from the hot spot, only the pre-shock zone
contributes appreciably to the radiation. However,
the fact that the pre-shock region reemits half of the
incoming short-wavelength radiation of the shock
(after the corresponding reprocessing) toward the
star is more important to us in this paper. To
take this effect into account, we did not modified
Lamzin's code but used the \texttt{CLOUDY} code (Ferland
et al. 1998), which was also applied by Calvet and
Gullbring (1998) in their calculations.

  The \texttt{CLOUDY} code is designed to compute the
thermal structure and radiation spectrum of a gas
layer on which radiation with a given spectrum is
incident from the outside. We used the version of
the code for a plane-parallel layer with solar elemental
abundances by assuming the external radiation to be
the sum of the radiations from the post-shock zone
and the hot spot. For simplicity, we assumed that
the spot radiated as a black body with an effective
temperature $T_{sp}$ defined by the relation
$$
\sigma T_{sp}^4 = \sigma T_{eff}^4 + {\rho_0 V_0^3\over 4}.
$$
We took the radiation spectrum of the post-shock
zone from Lamzin (1998) for the same values of $V_0$
and $N_0$ that were used to calculate the hot-spot
structure.

The radiation intensity of the post-shock zone was
assumed to be the same in all directions. For the
intensity of the radiation entering the pre-shock zone,
we can then write
$$
I_\nu = \frac{F_{ac}}{\pi} + B_\nu \left(T_{sp}\right).
$$
Note that within the isotropic approximation we
slightly underestimate the radiation density in the layers
adjacent to the shock front while simultaneously
overestimating it in more distant layers.

Not the intensity of the external radiation $I_\nu$ but
its value averaged over the solid angle $4\,\pi$ serves as an
input parameter of the \texttt{CLOUDY} code:
$$
4 \pi J_\nu = \int_{4\pi}{ I_\nu }\,{\rm d}\Omega =
2\pi\int\limits_0^1{ \left[\frac{F_{ac}}{\pi} +
B_{\nu}\left(T_{sp}\right)\right]} {\rm} {\rm d} \mu =
2F_{ac} + 2\pi B_\nu\left(T_{sp}\right).
$$
We took the distance from the shock front to the
point at which the gas temperature drops from its
maximum value to 6000~K as the layer thickness $H.$
At an expected diameter of the accretion-column
cross section $\gtrsim 3\times
10^{10}$ cm (a filling factor
$f \gtrsim 0.03$), the layer cannot be considered as a plane-parallel
one at a thickness $H \gtrsim 10^{10}$~cm. We limited
the thickness of the layer by this value even if the
temperature at its outer boundary exceeded 6000~K
in this case. Depending on the model, the layer has a
thickness $H=5{\times}10^9-10^{10}$~cm.

 In the \texttt{CLOUDY} code, it is assumed that the irradiated
gas layer is stationary, while the pre-shock
gas falls to the star with a velocity $V_0\sim 300$ km\,s$^{-1}.$
Since the thermal and ionization equilibrium is established
in a finite time, the temperature distributions in
the stationary and moving gases must differ (for more
detail, see Lamzin 1998). However, our checking
showed that these differences had virtually no effect
on the final result -- the thermal structure of the hot
spot.

   To calculate the heating of the stellar atmosphere,
we should determine the part of the radiation from the
pre-shock zone that is directed to the star. Assuming
that this radiation is isotropic, we have the following
relation between its intensity $I^{pre}_\nu(\mu)$ and the quantity
$4\pi J_\nu$ calculated by the \texttt{CLOUDY} code:
$$
I^{pre}_\nu(\mu) = {4\pi J_\nu \over 2\pi}.
$$

  The hot-spot radiation passes through the pre-shock
region on its way to the observer. The resulting
spectrum is the sum of the intrinsic radiation from the
region and the spot radiation partially absorbed in it:
$$
I_\nu(\mu) = I_\nu^{spot}(\mu)\,e^{-\frac{\tau_\nu}{\mu}} +
\int\limits_0^{\tau_\nu}
{S_{\nu}(t)\,e^{\frac{t-\tau_{\nu}}{\mu}}}\, {\rm d}t,
$$
where $I_\nu^{spot}(\mu)$ is the radiation intensity emergent
from the hot spot, $\tau_\nu$ is the optical depth of the
pre-shock zone at $\mu=1,$ and $S_\nu (\tau_\nu)$ is the source
function in the pre-shock zone computed with the
\texttt{CLOUDY} code only for the continuum.

   At small $\mu,$ the plane-parallel approximation
ceases to hold due to the finite cross section of the
accretion column. The value of $\mu_{||}$ at which about
half of the spot is observed through the side wall can
be estimated from the formula
$$
\mu_{||} =
\left[ 1+\frac{2f\left(1-2f\right)}{\left(H/R_*\right)^2}\right]^{-0.5},
$$
where $H/R_*$ is the ratio of the thickness of the pre-shock
zone to the CTTS radius, whose typical value
is $\sim 10^{11}$~cm. In our models, the plane-parallel approximation
holds for $\mu \gtrsim 0.5$ at $f \sim 0.03.$

  \end{document}